\documentclass[12pt]{article}
\usepackage{epsfig}

\def\beq{\begin{equation}}
\def\eeq#1{\label{#1}\end{equation}}
\def\eeqn{\end{equation}}

\def\beqa{\begin{eqnarray}}
\def\eeqa#1{\label{#1}\end{eqnarray}}
\def\eeqan{\end{eqnarray}}

\let\bar=\overbar

\def\Dslash{\not{\hbox{\kern-4pt $D$}}}
\def\dslash{\not{\hbox{\kern-2pt $\del$}}}

\def\msb{{\bar{\ssstyle M \kern -1pt S}}}

\def\Title#1{\begin{center} {\Large {\bf #1} } \end{center}}

\begin{document}

\Title{Gauge invariant nonlocal mass operator in YM theories}
\vspace{-3cm} \hfill LTH--768 \vspace{3.2cm}

\bigskip\bigskip


\begin{raggedright}

{\it D. Dudal\footnote{david.dudal@ugent.be}, N. Vandersickel, H. Verschelde\\
Ghent University, Department of Mathematical Physics and Astronomy,
Krijgslaan 281 S9, Belgium}
\bigskip\\
\it{J.A. Gracey\\
Theoretical Physics Division, Department of Mathematical Sciences,
University of Liverpool, P.O. Box 147, Liverpool, L69 3BX, United
Kingdom}
\bigskip\\
\it{M.A.L. Capri, V.E.R. Lemes, S.P. Sorella\\
Instituto de F\'{\i }sica, UERJ, Universidade do Estado do Rio de Janeiro, Rua S{\~a}o Francisco Xavier 524, 20550-013 Maracan{\~a}, Rio de Janeiro, Brasil} \bigskip\\
\it{R.F. Sobreiro\\CBPF, Centro Brasileiro de Pesquisas
F{\'\i}sicas, Rua Xavier Sigaud 150, 22290-180, Urca, Rio de
Janeiro, Brasil}

\end{raggedright}

\section{Introduction}
Lattice simulations in the Landau or maximal Abelian gauges have
made clear the need for mass parameters in the fitting functions
\cite{Langfeld:2001cz,Amemiya:1998jz,Bornyakov:2003ee}. These fits
involves e.g. a Yukawa propagator $\sim \frac{1}{q^2+M^2}$.
Moreover, phenomenological studies frequently make use of an
effective gluon mass \cite{Field:2001iu}. Theoretical studies, based
on Schwinger-Dyson equations and the Pinch Technique report a
dynamical gluon mass
\cite{Cornwall:1981zr,Aguilar:2004sw,Aguilar:2006gr}. Also
alternative analytical calculations gave such evidence, see e.g.
\cite{Verschelde:2001ia}. Finally, let us mention the issue of
$\frac{1}{q^2}$ power corrections in physical correlators, tackled
with QCD sum rules \cite{Chetyrkin:1998yr}, lattice and OPE
techniques \cite{Boucaud:2001st} or even via the AdS/QCD picture
\cite{Andreev:2006vy}. A natural question arising is where does such
a mass scale originate from? We recall that the standard Yang-Mills
action cannot contain a mass due to gauge invariance. The Higgs
mechanism is not a solution here, due to the associated gauge
symmetry breaking. A natural answer is that a dynamical mass scale
is generated by nonperturbative effects, in the form of a dimension
2 condensate. Of course, then the question pops up which $d=2$
operator $\mathcal{O}$ to consider? In our opinion, a few
requirements are to be met: $\mathcal{O}$ should be gauge invariant,
as it is supposed to enter physical quantities. It should be local,
as nonlocal actions are hard to handle/interpret, and it should be
renormalizable, as we want to perform quantum calculations, thus we
want consistent finite results, renormalization group (RG)
invariance, $\ldots$. Zakharov et al proposed to use $\displaystyle
A^2_{\min}=(VT)^{-1}\min_{U\in SU(N)}\int
d^4x\left(A_\mu^U\right)^2$ which is gauge invariant but nonlocal
 \cite{Gubarev:2000nz}. Only in the Landau gauge do we have $ \langle A^2_{\min}\rangle=A^2$
which is a renormalizable and  condensing local composite operator,
giving rise to effective dynamical
 gluon mass \cite{Verschelde:2001ia,Dudal:2002pq}. We could also use the mass operator based on the transverse gluon field: $(A_\mu^T)^2=\left[\left(\delta_{\mu\nu}-\frac{\partial_\mu\partial_\nu}{\partial^2}\right)A_\nu\right]^2$,
  or the non-Abelian Stueckelberg term, $ \mathcal{O}_{S}=\left( A_{\mu
}-\frac{i}{g}U^{-1}\partial _{\mu }U\right) ^{2}\;,\quad U=e^{ig\phi
^{a}T^{a}}$. The previous proposals are all classically equivalent
\cite{Capri:2005dy}, but face
 problems with nonlocality, nonpolynomiality, nonrenormalizability,
 $\ldots$. We proposed to study  $\mathcal{O}= \displaystyle \int d^4x F_{\mu\nu}^a
    \left[(D_\sigma^2)^{-1}\right]^{ab}F_{\mu\nu}^b$
    \cite{Capri:2005dy},
which we couple to the Euclidean YM action via
\begin{equation}
S_{\mathcal{O}}=-\frac{m^{2}}{4}\int d^{4}xF_{\mu \nu }^{a}\left[
\left( D_\sigma^{2}\right) ^{-1}\right] ^{ab}F_{\mu \nu }^{b}\;,
\label{massop}
\end{equation}
where $D_\sigma$ denotes the covariant derivative. This operator
{\cal O} already found use in 3D YM \cite{Jackiw:1997jga}. This
operator is clearly gauge invariant. Furthermore, it can be put
quite easily in a local form, since we can replace $S_\mathcal{O}$
with \footnotesize
\begin{eqnarray}
S_{\mathcal{O}}^\prime&=&  \frac{1}{4}\int d^{4}x\overline{B}_{\mu
\nu }^{a}D_{\sigma
}^{ab}D_{\sigma }^{bc}B_{\mu \nu }^{c}+\frac{1}{4}\int {%
d^{4}x}\overline{G}_{\mu \nu }^{a}D_{\sigma }^{ab}D_{\sigma
}^{bc}G_{\mu
\nu }^{c}+\frac{im}{4}\int d^{4}x\left( B-\overline{B%
}\right) _{\mu \nu }^{a}F_{\mu \nu }^{a}\,,
\end{eqnarray}\normalsize
where $B_{\mu\nu}^a,\overline{B}_{\mu\nu}^a$ are antisymmetric
bosonic and $G_{\mu\nu}^a,\overline{G}_{\mu\nu}^a$  fermionic
(ghost) fields in the adjoint representation. Notice that for $m=0$,
we have in fact introduced a unity.
\section{Renormalization analysis}
We continue our investigation from the starting action\footnotesize
\begin{eqnarray}\label{start}
S&=&\underbrace{\frac{1}{4}\int d^{4}xF_{\mu \nu }^{a}F_{\mu \nu }^{a}}_{S_{YM}}+\underbrace{\int d^{4}x\;\left( \frac{\alpha }{2}b^{a}b^{a}+b^{a}%
\partial _{\mu }A_{\mu }^{a}+\overline{c}^{a}\partial _{\mu }D_{\mu
}^{ab}c^{b}\right)}_{S_{gf}} +\frac{1}{4}\int
d^{4}x\overline{B}_{\mu \nu }^{a}D_{\sigma
}^{ab}D_{\sigma }^{bc}B_{\mu \nu }^{c}\nonumber\\&+&\frac{1}{4}\int {%
d^{4}x}\overline{G}_{\mu \nu }^{a}D_{\sigma }^{ab}D_{\sigma
}^{bc}G_{\mu
\nu }^{c}+\frac{im}{4}\int d^{4}x\left( B-\overline{B%
}\right) _{\mu \nu }^{a}F_{\mu \nu }^{a}\,.
\end{eqnarray}
\normalsize As the reader will notice, we imposed a linear gauge
fixing, encoded in  $S_{gf}$. The complete action enjoys a nilpotent
$BRST_1$ symmetry, generated by \footnotesize
\begin{eqnarray}\label{brst1}
s_1A_{\mu }^{a} &=&-D_{\mu }^{ab}c^{b}\;,\quad s_1c^{a}
=\frac{g}{2}f^{abc}c^{b}c^{c}\;,  \quad \ s_1B_{\mu \nu }^{a}
=gf^{abc}c^{b}B_{\mu \nu }^{c}\;,\quad s_1\overline{B}_{\mu \nu
}^{a} =gf^{abc}c^{b}\overline{B}_{\mu \nu }^{c}\;,
\nonumber\\
s_1G_{\mu \nu }^{a} &=&gf^{abc}c^{b}G_{\mu \nu }^{c}\;, \quad
s_1\overline{G}_{\mu \nu }^{a} =gf^{abc}c^{b}\overline{G}_{\mu \nu
}^{c}\;,\quad s_1\overline{c}^{a} =b^{a}\;,  \quad s_1b^{a}
=0\;,\quad s_1^2=0\,.
\end{eqnarray}
\normalsize For $m=0$, we can identify a nilpotent ``supersymmetry''
$\delta_s$ \cite{Capri:2006ne},
\begin{eqnarray}
\delta_s B_{\mu\nu}^a &=& G_{\mu\nu}^a\;,\quad \delta_s
\overline{G}_{\mu\nu}^a = \overline{B}_{\mu\nu}^a\;,\quad \delta_s
\Psi=0 \textrm{ for all other fields }\Psi\;,\quad \delta_s^2=0\,,
\end{eqnarray}
which can be invoked to establish another nilpotent $BRST_2$
invariance for $m=0$,
\begin{equation}
    s_2=s_1+\delta_s\,,\quad s_2^2=0\,.
\end{equation}
For $m\neq0$, we can embed the action (\ref{start}) into a
``larger'' model with external sources \cite{Capri:2005dy}. In a
particular physical limit, the original model is recovered. The
reason why we introduced the ``larger'' model is because it
simplifies the renormalizability analysis. If we can prove the
renormalizability of this action, then also the physically relevant
action will be renormalizable as a special case. The extended model
obeys many Ward identities, including a Slavnov-Taylor identity. We
constructed the most general action compatible with these Ward
identities, and when the smoke cleared after taking the physical
limit, we arrived at a renormalizable local action with mass terms
\cite{Capri:2006ne}, 
\begin{eqnarray}
  S_{phys} &=& S_{cl} +S_{gf}\;,\label{completeaction}\nonumber\\
  S_{cl}&=&\int d^4x\left[\frac{1}{4}F_{\mu \nu }^{a}F_{\mu \nu }^{a}+\frac{im}{4}(B-\overline{B})_{\mu\nu}^aF_{\mu\nu}^a
  \right.\nonumber\\&+&\left.\frac{1}{4}\left( \overline{B}_{\mu \nu
}^{a}D_{\sigma }^{ab}D_{\sigma }^{bc}B_{\mu \nu
}^{c}-\overline{G}_{\mu \nu }^{a}D_{\sigma }^{ab}D_{\sigma
}^{bc}G_{\mu \nu
}^{c}\right)\right.\nonumber\\
&-&\left.\frac{3}{8}%
m^{2}\lambda _{1}\left( \overline{B}_{\mu \nu }^{a}B_{\mu \nu
}^{a}-\overline{G}_{\mu \nu }^{a}G_{\mu \nu }^{a}\right)
+m^{2}\frac{\lambda _{3}}{32}\left( \overline{B}_{\mu \nu
}^{a}-B_{\mu \nu }^{a}\right) ^{2}\right.\nonumber\\&+&\left.
\frac{\lambda^{abcd}}{16}\left( \overline{B}_{\mu\nu}^{a}B_{\mu\nu}^{b}-\overline{G}_{\mu\nu}^{a}G_{\mu\nu}^{b}%
\right)\left( \overline{B}_{\rho\sigma}^{c}B_{\rho\sigma}^{d}-\overline{G}_{\rho\sigma}^{c}G_{\rho\sigma}^{d}%
\right) \right]\,,
\end{eqnarray}
which is still $BRST_1$ invariant (but not $BRST_2$!), whereby the
classical part $S_{cl}$ is gauge invariant. For reasons of
renormalizability, we had to introduce the scalar (mass) couplings
$\lambda_{1,3}$ and the gauge invariant tensor coupling
$\lambda^{abcd}$. Notice that the new quartic interaction
$\propto\lambda^{abcd}$ in the novel fields spoil the unity. One
might therefore question the equivalence with massless YM theories
when $m=0$. In \cite{Capri:2006ne}, we have been able to show that
\begin{equation}
    \left\langle \mbox{YM functional}\right\rangle_{S_{YM}+S_{gf}}= \left\langle \mbox{YM functional}\right\rangle_{S_{phys}+S_{gf}}
\end{equation}
by making use of the $\delta_s$-cohomology, meaning that expectation
values of ``pure'' YM functionals remain unchanged when calculated
in our massless model. As a consequence, there will be no
$\lambda^{abcd}$ independence in those pure YM Green functions, and
RG functions of the original YM quantities will remain unchanged, as
long as massless renormalization schemes are employed. This fact was
confirmed by explicit 1- and 2-loop computations in
\cite{Capri:2005dy,Capri:2006ne}. In addition, also several other RG
functions were computed. The obtained results were consistent with
the Ward identities and confirmed the renormalizability explicitly.
We end this section by mentioning that although the $BRST_2$
symmetry is broken, with an associated broken Slavnov-Taylor
identity $ST_2$, it is nevertheless possible to show that the
ensuing Ward identities between a large class of Green functions are
as if the $ST_2$ would be unbroken \cite{Capri:2007ix}.

\section{Unitarity analysis}
Since we have a massive gauge model with nilpotent $BRST_1$ symmetry
generator, we might hope that the theory would be unitary. Assuming
that we start from the action, and that we take the elementary field
excitations as asymptotic states, how can we prove the unitarity of
the ${\cal S}$-matrix? This rather complicated task was performed in
\cite{Dudal:2007ch}, whereto we refer for all details. One of the
main tools was the use of the (free) BRST cohomology. We start from
the free action $S_0$ with free BRST symmetry $s_0$ with nilpotent
charge ${\cal Q}_0$ (${\cal Q}_0^2=0$). Physical states
$\vert\psi_p\rangle$ are defined as belonging to the ${\cal
Q}_0$-cohomology, i.e.
\begin{equation}\label{1}
    \vert\psi_p\rangle\in\mathcal{H}_{phys}\Leftrightarrow\mathcal{Q}_0\vert\psi_p\rangle=0\;,\quad\vert\psi_p\rangle\neq
\vert\ldots\rangle+\mathcal{Q}_0\vert\ldots\rangle\,,\quad {\cal
Q}_0\vert\ldots\rangle=0\,.
\end{equation}
Then the 2 remaining questions are: is this definition invariant
under time evolution, described by the $\mathcal{S}$-matrix, and do
the physical states have a positive norm, a conditio sine qua non
for a sensible quantum theory. Concerning the time evolution, we
recall that in the operator language
$\mathcal{S}=\mathcal{T}\left[e^{-i\int_{-\infty}^{+\infty}H_{int}(t)dt}\right]
$, so that $\mathcal{S}\vert\psi_p\rangle\in\mathcal{H}_{phys}$, if
we require that $\left[\mathcal{S},\mathcal{Q}_0\right]=0$. We can
pass to the path integral language to rephrase this condition into
one for the action $S$, namely
\begin{eqnarray}
s_0 e^{iS}&\cong&0\quad\mbox{on-shell, i.e. modulo free equations of
motion}\,.
\end{eqnarray}
This equation can then be solved iteratively, in particular order by
order in the available coupling constant(s).
\begin{eqnarray*}
\mbox{action}\,\, S&=&S_0+S_1+\ldots\;,\quad \mbox{BRST}\,\,
s=s_0+s_1+\ldots\,,
\end{eqnarray*}
while it is proven that
\begin{eqnarray}
    (s_0+\ldots +s_i) (S_0+\ldots+S_i)&=&0\,,\quad (s_0+\ldots+s_i)^2=0\; \mbox{to $i^{th}$
    order}\,.
\end{eqnarray}
If we are lucky, the procedure stops at finite order, and we end up
with an action $S$, invariant under a BRST symmetry $s$ with
nilpotent generator ${\cal Q}$, with the desired property that the
physical subspace $\mathcal{H}_{\mbox{\tiny phys}}$ is invariant
under time evolution. We used a ``backward'' argument: if we start
from an action $S$ with nilpotent symmetry charge ${\cal Q}$, then
this is a solution of the previous procedure starting from the
corresponding free counterparts $S_0$ and ${\cal Q}_0$, obtained by
switching of any couplings. Having answered the time evolution
question, we are still left with the positive norm issue. We suffice
by saying this required a rather lengthy and technical study. One of
the problems faced was the occurrence of multipole fields.
Nevertheless, a complete analysis was provided in
\cite{Dudal:2007ch}, yielding a negative result: negative norm
states do appear in the physical subsector of the theory, therefore
it is not unitary. The reader might have immediately questioned the
wisdom of trying to prove the unitarity of this model, as it has its
root in a nonlocal field theory, which are known to have problems
with ghost states. Next to the presence of a nilpotent BRST
invariance, there is however another reason why this endeavour was
worth the effort. Specializing to the Abelian case, it can be shown
that after integrating out the auxiliary fields, we obtain the
Abelian Stueckelberg model, where the Stueckelberg scalar has been
integrated out. We recall that the Abelian Stueckelberg model is
renormalizable and unitary, see e.g. \cite{Ruegg:2003ps} for a
review. If we would analyze the unitarity of the Abelian version of
our model, we would run into exactly the same problem as in the
non-Abelian case, i.e. the presence of negative norm states in the
physical subspace. Apparently, the way of localizing a nonlocal
action plays a substantial role.

\section{Discussion}
Since our action is perturbatively equivalent with YM in the
massless case, we could take it as starting point instead of YM. As
proven, we can couple mass terms to it, without ruining
renormalization/gauge invariance requirements. The next challenge
would of course be trying to construct a sensible gap equation to
produce a dynamically generated value for $m\propto \Lambda_{QCD}$.
Afterwards, we could start looking at potential nonperturbative
$\frac{m^2}{q^2}$ power corrections appearing in gauge (in)variant
correlators. We are thus interested in nonperturbative effects in an
asymptotically free theory, which occur in an energy region below
the high energy (asymptotic) region where the elementary fields
excitations are observables. Since this is YM ($m\equiv0$), there is
no unitarity issue at high energies: 2 transverse gluon
polarizations are physical. At lower energies, the gluons etc still
are our effective degrees of freedom, but will behave like quasi
particles, corrected by nonperturbative effects due to the generated
$m\neq0$. Lack of unitarity in terms of gluons is hence not a
problem, but rather finding the correct physical degrees of freedom,
which of course corresponds to the task of proving confinement.

\section*{Acknowledgments.}
D.~Dudal is a Postdoctoral Fellow and N.~Vandersickel a PhD Fellow
of the Research Foundation-Flanders (FWO Vlaanderen). The Conselho
Nacional de Desenvolvimento Cient\'{\i}fico e Tecnol\'{o}gico
(CNPq-Brazil), the SR2-UERJ and the Coordena{\c{c}}{\~{a}}o de Aperfei{\c{c}}%
oamento de Pessoal de N{\'\i}vel Superior (CAPES) are gratefully
acknowledged for financial support.

\end{document}